# Energy and Atomic Mass Dependence of Nuclear Stopping Power in Relativistic Heavy-Ion Collisions in Interacting Gluon Model

Q. J. Liu[1*], W. Q. Chao[1,2†] and G. Wilk[3‡]

[1]*Institute of High Energy Physics, Academia Sinica,*
*P.O. Box 918(4), 100039, Beijing, P. R. China*
[2]*CCAST (Word Lab.), P.O. Box 8730, 100080, Beijing, P. R. China*
[3]*Soltan Institute for Nuclear Studies, Nuclear Theory Department,*
*Warsaw, Poland*

## Abstract

We present a Monte-Carlo simulation of energy deposition process in relativistic heavy-ion collisions based on a new realization of the Interacting-Gluon-Model (IGM) for high energy $N-N$ collisions. In particular we show results for proton spectra from collisions of $E_{lab} = 200~GeV/N$ $^{32}$S beam incident on $^{32}$S target and analyze the energy and mass dependence of nuclear stopping power predicted by our model. Theoretical predictions for proton rapidity distributions of both $^{208}$Pb + $^{208}$Pb collisions at $E_{lab} = 160~GeV/N$ CERN SPS and $^{197}$Au + $^{197}$Au at $\sqrt{s_{NN}} = 200~GeV$ BNL RHIC are given.

## 1 Introduction

Whether the central rapidity region is baryon free or not after the collision is an important issue in the field of relativistic heavy-ion collisions. These two different pictures imply different mechanisms for QGP formation, if QGP is really produced there, and thus will make the detection of QGP different for the two situations. The subject is directly related to how the C.M.S energies of the colliding nuclei are deposited in this region, i.e., to the nuclear stopping power. Though much work that has been done on both theoretical and experimental aspects of this subject indicates so far a baryon rich picture (i.e., complete stopping) [1] for heavy systems such as gold on gold below the AGS energies, where initial conditions for QGP formation are not well fulfilled, a common consensus on the issue has not yet been reached for heavy-ion collisions at CERN SPS to BNL RHIC and CERN LHC energies, where one expects the formation of QGP. For example, FRITIOF [2] predicts fully baryon free in the central rapidity area for collisions of gold on gold at RHIC energies while VENUS [3] and DTU [4]) predict the opposite, although except for this, all models have almost the same predicting abilities from AGS to SPS energies.

---
*e-mail: liuqj@bepc3.ihep.ac.cn
†e-mail: zhaowq@bepc3.ihep.ac.cn
‡e-mail: wilk@fuw.edu.pl

Whereas in the above mentioned models the main degrees of freedom are quarks (valence in LUND and with the sea-quark admixture in VENUS and DTU) and gluons are treated collectively as non-interacting strings, the Interacting-Gluon-Model (IGM) [5, 6, 7] pays more attention to the role of gluons in high energy collisions, following the fact that in the perturbative QCD (PQCD) gluon-gluon interactions are much more important than those between quarks [8]. IGM was already applied successfully to nuclear collisions, however either in the *coherent nuclear tube* version which does not allow for calculation of the nucleonic spectra [6] or in the version devoted to the specific problem of looking for *coherent* vs *chaotic* aspects of nuclear collisions, also not directly suitable for our present purposes [7].

As both Ref.[6] and Ref.[7] have already shown the basic ability of IGM to describe adequately essential details of multiparticle production processes, we shall concentrate in this letter only on the study of nuclear stopping power provided by IGM when extended to heavy-ion collisions at C.M.S energies no smaller than those of SPS.

## 2 Brief review on the IGM for high energy $N-N$ process

The IGM can be summarized as follows [5]: According to PQCD: $\sigma_{gg} \gg \sigma_{gq} \gg \sigma_{qq}$, where $\sigma_{gg}$, $\sigma_{gq}$ and $\sigma_{qq}$ are the interaction cross sections of a gluon-gluon, gluon-quark and quark-quark pair, respectively. Therefore, assuming that the above inequality holds also in soft interactions, in each event one has he following picture[5]:

- Two colliding hadrons are represented by the valence quarks carrying their quantum numbers plus the accompanying clouds of gluons (which also include the sea $q\bar{q}$ pairs and should therefore be regarded as effective ones). In the course of the collision, gluonic clouds interact strongly and form *mini-fireballs* (MF's) which are supposed to populate mainly in the central rapidity region of the reaction.

- The valence quarks (together with the gluons which did not interact) get excited and form *leading jets* (LJ's) (or *beam jets*) which are supposed to populate mainly in the fragmentation regions of the reaction. Here we shall assume further that LJ's are identical with leading particles (practically nucleons).

The above mentioned MF's eventually form a lump of gluonic matter which is called a central fireball (CF). It is this CF that governs the final state particle production in the central rapidity region.

In the framework of the IGM, if fractions of $x$ and $y$ of the energy momentum of the incoming nucleons are deposited in central rapidity region and form a CF, then the energy $E$ and momentum $P$ and invariant mass $M$ and rapidity $\Delta$ of this CF are expressed as (all masses are neglected here):

$$E = \frac{1}{2}\sqrt{s}\ (x+y), \qquad P = \frac{1}{2}\sqrt{s}\ (x-y) \qquad (1)$$

and

$$M = \sqrt{xys}, \qquad \Delta = \frac{1}{2}\ \ln(\frac{x}{y}). \qquad (2)$$

while the probability distribution for producing such a CF is [6, 11]

$$\chi(x,y) = \chi_0 \ \exp[-\frac{(x - \langle x \rangle)^2}{2\langle x^2 \rangle} - \frac{(y - \langle y \rangle)^2}{2\langle y^2 \rangle}] \ \theta(xy - K_{min}^2) \qquad (3)$$

where $\chi_0$ is a normalization factor and $K_{min} = M_0/\sqrt{s}$ stands for the minimal inelasticity allowed. There is simple relation among the moments $\langle x \rangle$, $\langle x^2 \rangle$ and $\langle y \rangle$, $\langle y^2 \rangle$, which is valid at high energy, namely

$$\langle x \rangle \simeq 2\langle x^2 \rangle \simeq \langle y \rangle \simeq 2\langle y^2 \rangle = \frac{\alpha p_h^2}{M_0^2 \sigma_{in}^{NN}(s)} \qquad (4)$$

where $\alpha$ is a constant related to the gluonic cross-section, $p_h$ represents fraction of the hadronic energy-momentum attributed to gluons in hadron structure function [9] (in fact we have here $p_h^2 = p_{projectile} \cdot p_{target} = p^P \cdot p^T$), $M_0$ is the lightest mass of the possible CF and $\sigma_{in}^{NN}(s)$ is the inelastic cross section for high energy $N - N$ process.

We have thus essentially three input parameters (although highly correlated, cf. Eq.(4)): $M_0$ delineating the gluonic phase space (i.e., $x\ y \geq K_{min}^2 = M_0^2/s$), $\alpha$ fixing the strength of gluonic interaction and product of $p_h$'s dictating what part of energy-momentum of initial nucleons is used to form CF. [1]

## 3 IGM and high energy heavy-ion collisions

We proceed now to simulation of energy deposition process in relativistic collisions of two nuclei $A$ and $B$ (with nucleus $B$ incident on nucleus $A$), where $A$ and $B$ denote also their corresponding atomic numbers. In IGM it is realized in steps according to the following algorithm [13]:

**(i)** We simulate positions of all the $A$ and $B$ nucleons, in the rest frames of colliding nuclei with the respective coordinate origins located at their mass centers and according to their nucleon densities $\rho_{A,B}(\mathbf{r})$ given by the usual three-parameter Woods-Saxon form [14].

**(ii)** For every heavy-ion collision event we sample the impact parameter $\mathbf{b}$ using the Glauber model [15], in which the inelastic cross section of high energy $A - B$ process is given by

$$\sigma_{in}^{AB} = \int d\mathbf{b}[\ 1 - e^{-\sigma_{in}^{NN} T_{AB}(\mathbf{b})}] \qquad (5)$$

with

$$T_{AB}(\mathbf{b}) = \int d\mathbf{b_A}\ dz_A\ d\mathbf{b_B}\ dz_B\ \rho_A(\mathbf{b_A}, z_A)\ \rho_B(\mathbf{b_B}, z_B)\ \delta(\mathbf{b} - \mathbf{b_A} + \mathbf{b_B}) \qquad (6)$$

being the thickness function normalized to 1. Consequently, the probability distribution for the impact parameter $\mathbf{b}$ is given by

$$\frac{dP(\mathbf{b})}{d\mathbf{b}} = \frac{1 - e^{-\sigma_{in}^{NN} T_{AB}(\mathbf{b})}}{\sigma_{in}^{AB}}. \qquad (7)$$

---

[1] It is perhaps worth mentioning that such a formulated IGM was, in addition to Refs.[[5, 6, 7]], used also to discuss the problem of inelasticity in cosmic ray experiments [10], to connect the limited number of clans in multiparticle distributions with the limited behaviour of inelasticity [11], and recently it was also discussed in connection with the possible mini-jets production [12].

Once the impact parameter **b** is selected according to Eq.(7), the transverse and longitudinal coordinates of all colliding nucleons can be determined from their initial coordinates. Finally all nucleons in both colliding nuclei are ordered according to their z-coordinates in lab. frame, which results in: $z_P(B) < z_P(B-1) < z_P(B-2) < \cdots < z_P(1) < z_T(1) < z_T(2) < \cdots < z_T(A)$ ($z_P(i)$ denotes the z-coordinate of the $i$th nucleon in nucleus $B$ with $i \in (1, B)$ and $z_T(j)$ stands for the z-coordinate of $j$th nucleon in nucleus $A$ with $j \in (1, A)$).

(iii) Finally, we simulate the nucleus-nucleus collision as proceeding via a number of consecutive $N - N$ like collisions in the following way:

- For a pair of nucleons $ij$: $i = 1, \ldots, B$ from nucleus $B$ and $j = 1, \ldots A$ from nucleus $A$, we check first if they interact. They do it if their transverse distance is smaller than $\sqrt{\frac{\sigma_{tot}^{NN}(s_k)}{\pi}}$ and $\sqrt{s_k} \geq 2.14\ GeV$. Otherwise this pair will not collide and the four-momenta of each of the two nucleons are kept unchanged. Here $\sqrt{s_k}$ stands for the energy of the $k$-th pair in its c.m. frame. Both impinging nucleons and the produced leading particles are assumed to propagate along straight lines.

- For the pair of nucleons which do interact, we carry out the corresponding nucleon-nucleon collision which proceed at current energy of the colliding nucleonic pair, $\sqrt{s_k}$. whether th collision is elastic or inelastic is sampled according to the ratio of the corresponding cross sections $\sigma_{el}^{NN}(s_k)$ and $\sigma_{tot}^{NN}(s_k)$ taken from experimental data [16]. For inelastic collision, the IGM is called for providing a new gluonic CF and two new leading particles formed from both participating nucleons. The energy-momentum conservation implies the continuous degradation of the collision energy of interacting nucleonic pairs along the chain of consecutive scatterings. As all the 3 essential parameters of the IGM are kept unchanged with $k$ (and equal to: $\alpha p_h^2 = 0.05\ GeV^2 fm^2$, $M_0 = 0.22\ GeV)^2$, the above procedure corresponds to a tacit assumption of restoration of the gluonic clouds in the nucleons propagating through the nuclear matter after every inelastic collision. The magnitudes of the corresponding transverse momenta of colliding nucleons are sampled according to a Gaussian distribution with $\langle p_T \rangle = 0.45\ GeV/c$ [17]. For elastic scattering, we use differential cross section from Ref.[18] to generate new four-momenta of colliding nucleons.

Such an implementation of IGM to nuclear collisions differs distinctly from that in Ref.[6] or Ref.[7]. In fact, it can be regarded as a new realization of the IGM, similar to the so called "model C"[19]. Contrary to Refs.[[6, 7]], it allows the presence of additional energy-momentum transfer from the valence quark component of the colliding nucleons to gluons during the nuclear collision process (specifically: in between the consecutive scatterings). As we shall see below, such a scenario leads to the nuclear stopping compatible with experimental data.

## 4  Results and discussion

Fig.1 and Fig.2 display the event-normalized proton rapidity distributions in central collisions (corresponding to zero impact parameter, $\mathbf{b} = 0$, in the above IGM.) of $^{32}$S + $^{32}$S and $^{197}$Au + $^{197}$Au, at CERN SPS and BNL RHIC energies, respectively. The noticeable feature of the displayed results is the increase of baryon number in the central rapidity region with increasing atomic mass of colliding nuclei and the decrease of baryon number in the region with growing energy.

---

[2] As mentioned before, $\alpha$ and $p_h$'s are strongly correlated and therefore usually given as product like in this case above. If we assume that like in Ref.[5], $p_h \sim 0.5$, then $\alpha = 0.2\ GeV^2 fm^2$.

A useful measure [20, 21] of the nuclear stoping power characterizing the baryon content of the central rapidity region is the mean rapidity shift $\langle \Delta y_p \rangle$ of the projectile participant protons from their original beam rapidity. The larger is this value, the stronger is the stopping of the colliding system and the more baryons are showing up in the central rapidity region. In order to compare this measurement to those of other colliding systems at different energies, a relative stopping parameter, which also is an alternative of the event-normalized proton rapidity distributions, was introduced as follows[22]:

$$S = \frac{\langle \Delta y_p \rangle}{y_{beam} - y_{cm}} \qquad (8)$$

with $y_{cm}$ being the rapidity of center of mass of all participant nucleons. The values of $S$ for collisions of two mass symmetric nuclei as calculated in our model at CERN SPS and BNL RHIC energies are tabulated in Table 1. They confirm the statements suggested already by Fig.1 and Fig.2, namely that heavy systems are more stopped than light ones at the same energy $\sqrt{s_{NN}}$ whereas for the same colliding system, the larger the $\sqrt{s_{NN}}$, the weaker is the nuclear stopping power.

Table 1: The energy and mass dependence of relative stopping parameter $S$.

| System | $\sqrt{s_{NN}}$ ($GeV$) | b(fm) | Theor. | Exp. |
|---|---|---|---|---|
| $^{32}$S + $^{32}$S | 20 | 0-1 | 0.51 | 0.53 |
| $^{32}$S + $^{32}$S | 20 | 0 | 0.49 | No |
| $^{197}$Au+$^{197}$Au | 20 | 0 | 0.67 | No |
| $^{32}$S + $^{32}$S | 200 | 0 | 0.35 | No |
| $^{197}$Au+$^{197}$Au | 200 | 0 | 0.59 | No |

The first conjecture results from the successive multiple scatterings of colliding nucleons and this effect manifested itself in recent experimental data [21] from NA35 Collaboration. The effect would increase the number of collisions an incident nucleon experiences, and would become stronger for heavier colliding systems. Therefore with *a priori* more inelastic collisions (which means more energy loss) present, heavy system are stopped more than light one at the same colliding energy. In fixed target $E_{lab} = 200~GeV/N$ $^{32}$S + $^{32}$S central collisions, there is good agreement between theoretical calculations and experimental results both for the relative stopping parameter $S$ as shown in Table.1 and for the proton spectra as shown in Fig.3 and Fig.4. This agreement indicates that the multiple scattering effect has been properly accounted for in our simulation.

The second statement is caused by the energy dependence of cross section $\sigma_{in}^{NN}(s)$ in elementary high energy $N - N$ processes. The smaller energy loss of a participant projectile proton or a weaker stopping power corresponds to a smaller value of $\langle x \rangle$, and vice versa. Therefore a larger energy $\sqrt{s_{NN}}$ of the colliding system leading to a larger $\sigma_{in}^{NN}(s)$ according to Ref.[16], would results in a smaller value of $\langle x \rangle$ according to Eq.(4): $\langle x \rangle \simeq \frac{\alpha p_h^2}{M_0^2 \sigma_{in}^{NN}(s)}$, and thus leads to a weaker nuclear stopping power.

As a theoretical prediction we shown in Fig.5 the event-normalized proton rapidity distributions in central collisions of $E_{lab} = 160~GeV/N$ $^{208}$Pb + $^{208}$Pb with $bfb = 0~fm$, for which we predict the relative stopping parameter $S = 0.67$.

The theoretical results concerning proton spectra shown in Fig.3 and Fig.4 are both normalized to the experimentally observed proton number in order to facilitate comparison with data. Because in this letter we have assumed for simplicity that all participant protons are observed in the final state of the heavy-ion collisions, our estimations provide only upper limits for proton rapidity distributions. However, in mass symmetric heavy-ion collisions, it is expected to result in a reasonable conclusion on whether or not the central rapidity region is baryon free. This happens because other baryon productions (mainly strange baryon productions) which contribute to the net baryon rapidity distributions, proceed via fragmentation of the excited forward (backward) baryonic strings, (cf., Ref.[3]), and thus manifested themselves mostly in the projectile (target) rapidity region (also their abundances are much less than that of the protons considered there).[3]

## 5 Conclusions

Extending the IGM for high energy $N - N$ process to relativistic heavy-ion collisions, we simulated energy deposition process in high energy nucleus-nucleus collisions. Our theoretical results of proton spectra in $E_{lab} = 200\ GeV/N$ $^{32}$S + $^{32}$S central collisions are in good agreement with experimental data. Moreover, according to our calculations, the amount of baryons present in the central rapidity region in relativistic heavy-ion collisions, is determined both by the mass and by the energy $\sqrt{s_{NN}}$ of the colliding systems. It decreases with the increase of $\sqrt{s_{NN}}$ (indicating a weaker nuclear stopping power) and dramatically increases with the mass of the colliding systems (the final state baryons are inclined to be piled up in the central rapidity region signifying a stronger nuclear stopping power). Finally, if QGP will really be produced in $E_{lab} = 160\ GeV/N$ $^{208}$Pb + $^{208}$Pb central collisions at SPS or in $\sqrt{s_{NN}} = 200\ GeV$ $^{197}$Au + $^{197}$Au central collisions at RHIC, it is predicted to be a baryon rich one, not a baryon free one. Although in this respect our conclusion is generally similar to that of the string models (cf., VENUS [3], DTU [4] and also [23]), we foresee a difference between IGM and the above mentioned models in the LHC energy range. The reason is the limited (i.e., at most increasing to a limited value) energy behaviour of inelasticity parameter of hadronic collisions predicted by IGM [10, 12] whereas it increases rather strongly in models presented in Refs.[[3, 4, 23]] (cf. Ref.[10] for more details). It means that *asymptotically* we should have generally *broader* distribution of baryons than those models predict, with width asymptotically constant (versus decreasing one in their case).

**Acknowledgement:** This work is partly supported by Academia Sinica and National Science Foundation of China.

---

[3]Some other papers [23, 24] which discuss strange baryon production from different point of view, also have difficulties with correct description of the experimental data [24, 25]. This is the current situations about the study of baryonic strange particle productions. Therefore, in our calculations, the baryon rapidity distributions from strange baryon production is supposed to have similar shape as that of protons based on measurements reported in Ref.[26] and Ref.[24].

# Figure captions

**Fig. 1** Theoretical results of event-normalized proton rapidity distributions for fixed target central collisions ($b = 0\ fm$) of $E_{lab} = 200\ GeV/N$ $^{32}$S + $^{32}$S (solid line) and $^{197}$Au + $^{197}$Au (dotted line).

**Fig. 2** Theoretical results of event-normalized proton rapidity distributions for central collisions ($b = 0\ fm$) of $^{32}$S + $^{32}$S (shown in solid line) and $^{197}$Au + $^{197}$Au (shown in dotted line) collisions at RHIC energy $\sqrt{s_{NN}} = 200\ GeV$.

**Fig. 3** Event-normalized proton rapidity distributions in $E_{lab} = 200\ GeV/N$ $^{32}$S + $^{32}$S fixed target central collisions. Solid line stands for theoretical result normalized to the experimentally observed number of protons. Experimental data are from reference [21].

**Fig. 4** Event-normalized proton transverse momentum distributions in fixed target central collisions of $^{32}$S + $^{32}$S with $E_{lab} = 200\ GeV/N$. Solid line stands for theoretical result normalized to the experimentally observed number of protons. Experimental data are from reference [21].

**Fig. 5** Theoretical prediction for event-normalized proton rapidity distributions in $E_{lab} = 160\ GeV/N$ and $b = 0\ fm$ fixed target $^{208}$Pb + $^{208}$Pb collisions.

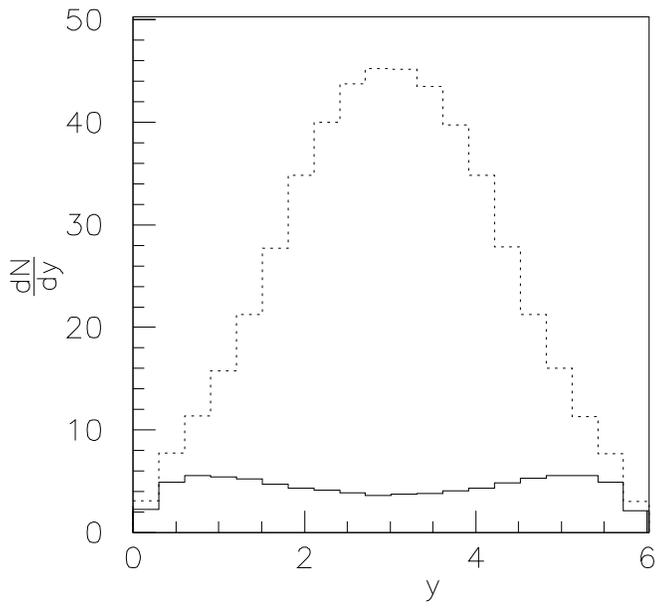

Fig.1

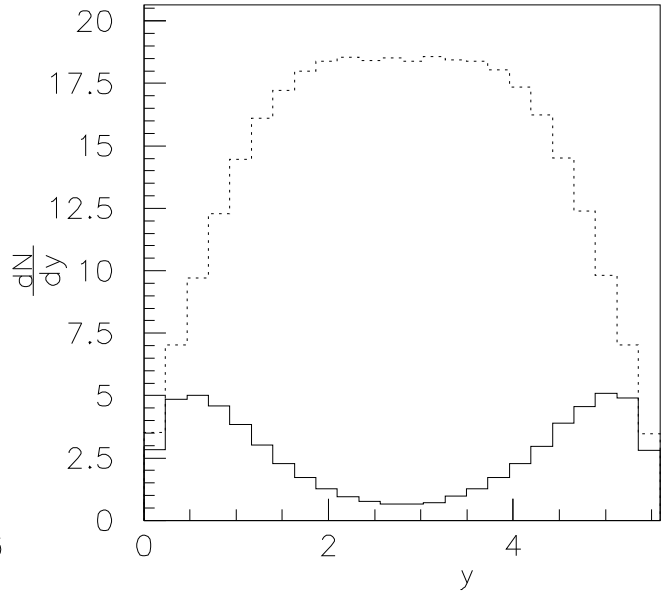

Fig.2

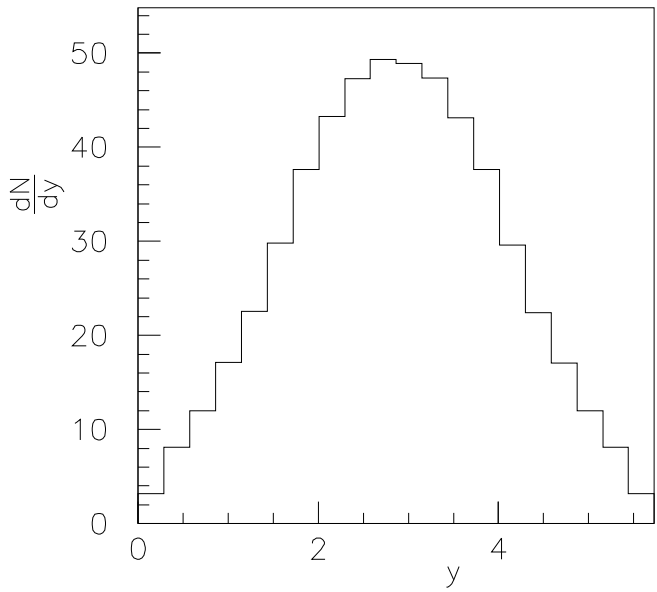

Fig.5

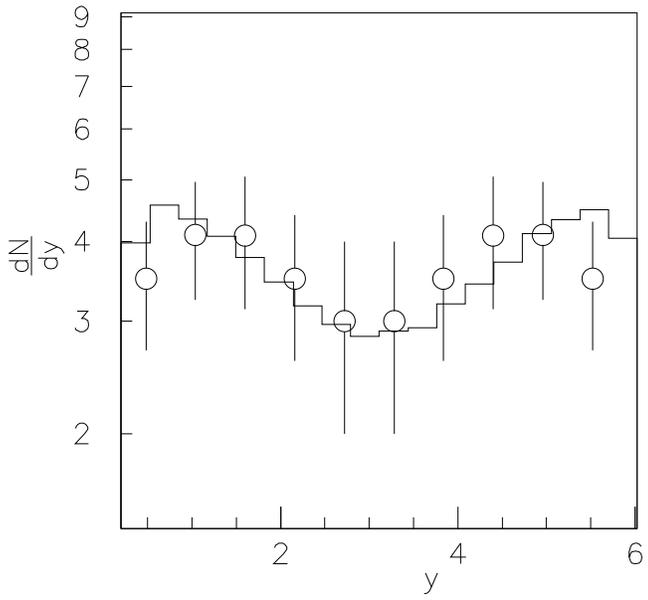

Fig.3

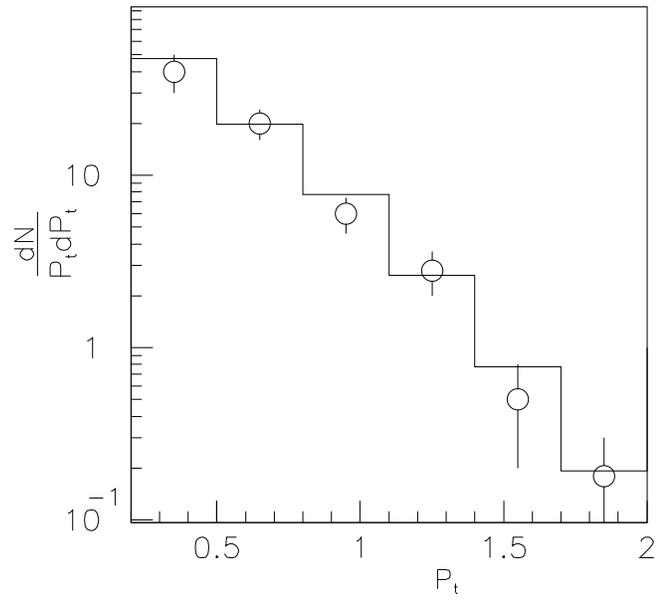

Fig.4